\documentclass[aps,prb,superscriptaddress,showpacs,floatfix]{revtex4-1}
\usepackage{amsfonts}
\usepackage{graphicx,amsmath,amssymb,xspace,epsfig,float,multirow,subfigure,tabularx}

\begin{document}

\title{Upper magnetic field in superconducting Dirac semi-metal}

\author{Baruch Rosenstein}
\email{baruchro@hotmail.com}
\affiliation{Electrophysics Department, National Chiao Tung University, Hsinchu 30050,
\textit{Taiwan, R. O. C}}
\author{B.Ya. Shapiro}
\email{shapib@mail.biu.ac.il}
\affiliation{Physics Department, Bar-Ilan University, 52900 Ramat-Gan, Israel}
\author{Dingping Li}
\email{lidp@pku.edu.cn}
\affiliation{School of Physics, Peking University, Beijing 100871, \textit{China}}
\affiliation{Collaborative Innovation Center of Quantum Matter, Beijing, China}
\date{\today}

\begin{abstract}
Temperature dependence of the upper critical field $H_{c2}$ of the Dirac
semi - metal (DSM) with phonon mediated pairing is considered within semi -
classical approximation. The low temperature dependence deviates from
conventional BCS superconductor with parabolic dispersion relation\cite{WHH}
even for large adiabaticity parameter, $\gamma =\mu /\left( \hbar \Omega
\right) $., where $\mu $ is the chemical potential and $\Omega $ - Debye
frequency. In particular the "reduced field", ratio of zero temperature $%
H_{c2}$ to derivative at critical temperature, $h^{\ast }=H_{c2}\left(
0\right) /\left( -T\frac{dH_{c2}}{dT}\right) |_{T_{c}}$, depends on $\gamma $
and can be extended beyond the adiabatic limit. The reduced magnetic field
ratio is universal (independent of the chemical potential, interaction
strength etc.) and smaller than the Werthamer ratio for clean
superconductors: $h^{\ast }=0.55$ for DSM, $h^{\ast }\left( 0\right) =0.69$
for parabolic band. The results are in good agreement compared with recent
experiments on $TaP$.
\end{abstract}

\maketitle

\section{Introduction}

Recently a large new class of 2D (including purely 2D materials like
graphene and surfaces of "topological insulators") and 3D multi-band
materials with qualitatively different band structure (Dirac point) near the
Fermi level was discovered. Unlike in conventional semi-metals with several
quasiparticle and hole bands, the Dirac points occur due to the band
inversion near the Fermi level. In many of them (sometimes under pressure)
superconductivity was observed at low temperatures \cite%
{Geim,superWSM2D,superWSM3D,Cao,PdTe2,Hf}. Dirac semi-metals (DSM) are
characterized by linear dispersion relation, and the chemical potential is
tunable and small. More importantly for the pairing\ of the quasiparticles
by phonons leading to "conventional" superconductivity is that their inter -
band tunneling is dominant \cite{frontiers,JCM15}.

In type I DSM, the band inversion results in Dirac points in low-energy
excitations being anisotropic massless "relativistic" fermions (Dirac cone
in dispersion relation, $\varepsilon =vp$). They exhibit several remarkable
properties like the chiral magnetic effect related to the chiral anomaly in
particle physics and tend to form the s-wave superconductivity \cite%
{DasSarma,FuBerg,Shapiro14,ruslan} of the second kind (recently discovered
type-II DSMs \cite{Goerbig1,Soluyanov} with tilted cones and nearly flat
bands tend to exhibit superconductivity of first kind \cite{PdTe2}).

Magnetic properties of the DSM turned superconductor of the second kind are
typically standard. The upper critical magnetic field $H_{c2}\left( T\right)
$ was measured in wide range of temperatures for different DSM and it
demonstrates behavior typical for vortex matter. In particular in layered
DSM (that possess relatively large Ginzburg number $Gi$) $MoTe_{2}$ \cite%
{MoTe2} the thermal fluctuations effects are observed in the vicinity of the
critical temperature $T_{c}$ \cite{Rosenstein18,strongfield, koshelev}. The
general theory based in Ginzburg - Landau approach (that is insensitive to
the microscopic details distinguishing between metal with parabolic bands,
DSM of high $T_{c}$ unconventional superconductors) is applicable for most
of the observed features.

However it was noticed very recently that the low temperature dependence of
the $H_{c2}\left( T\right) $ curve in some DSM, like \cite{ZrTe} $ZrTe_{5}$,
\cite{CdAs} $Cd_{3}As_{2}$ and\cite{TaP} $TaP$, significantly deviate from
that predicted by the quasi-classical microscopic theory for conventional
parabolic band superconductors \cite{Gorkov, WHH,Bulaevsky}. The low
temperature limit is beyond the universality range of the GL approach and
\textit{is} sensitive to the "microscopic details", especially when the
inter-band tunneling is present. The deviations were discussed in literature
for superconductors that have large gyromagnetic ratio\cite{Chandrasekhar},
but this is clearly not the case here. Therefore it is important to extend
the original Werhamer-Helfand-Hohenberg \cite{WHH} theory of the upper
critical field to the case of multi - band DSM with large tunneling between
the valleys \cite{koshelev,Rosenstein18,Rosenstein17}.

This is the purpose of the present work. The quasi-classic theory of the
phonon - mediated pairing in magnetic fields in DSM is developed in wide
range of temperatures and adiabatic parameters. In particular the reduced
magnetic field at zero temperature defined by $h^{\ast }=H_{c2}\left(
0\right) /\left( -TdH_{c2}/dT\right) |_{T_{c}}$ for different adiabaticity
ratio $\gamma =\mu /\hbar \Omega $ ($\mu $ is the chemical potential, while $%
\Omega $ is Debye frequency) is calculated. Even within the BCS limit $%
\gamma \rightarrow \infty $ the reduced magnetic field is smaller than the
conventional WHH \cite{WHH} (extended to the layered system by Bulaevsky
\cite{Bulaevsky} with the magnetic field directed perpendicular to the
layers) value $h_{WWH}^{\ast }=0.69$.

In addition, since in DSM the adiabaticity ratio in Weyl semi-metals is
relatively small, an important additional issue is the role of the
retardation effects of the phonon mediated pairing. Within the semiclassical
approach one can approach moderately adiabatic case, $\gamma \simeq 1$.

\section{Pairing in the DSM under magnetic field}

\subsection{Hamiltonian}

Dirac material typically possesses several sublattices. We exemplify the
effect of the DSM band structure on superconductivity using the simplest
model with just two sublattices denoted by $\alpha =1,2$. The effective
electron-electron attraction due to the electron - phonon coupling overcomes
the Coulomb repulsion and induces pairing. Typically in DSM there are
numerous bands. We assume that different valleys are paired independently
and drop all the valley indices (including chirality, multiplying the
density of states by $2N_{f}$). To simplify notations, we therefore consider
just one spinor (left, for definiteness), the following Weyl Hamiltonian\cite%
{JCM15}, \cite{Wang}.%
\begin{equation}
K=\int_{\mathbf{r}}\psi _{\alpha }^{s\dagger }\left( \mathbf{r}\right) \left
\{ -i\hbar v\left( D_{x}\sigma _{\alpha \beta }^{x}+D_{y}\sigma _{\alpha
\beta }^{y}\right) -i\hbar v_{z}D_{z}\sigma _{\alpha \beta }^{z}-\mu \delta
_{\alpha \beta }\right \} \psi _{\beta }^{s}\left( \mathbf{r}\right) .
\label{K}
\end{equation}%
Here $v$ is Fermi velocity assumed isotropic in the plane $x-y$
perpendicular to the applied magnetic field, $v_{z}$ is the velocity in
magnetic field direction. Chemical potential is denoted by $\mu $. Pauli
matrices $\sigma $ operate in the sublattice space (the indices $\alpha
,\beta $ will be termed the pseudo-spin projections) and $s$ is spin
projection. Magnetic field appears in the covariant derivatives via the
vector potential, $D_{i}=\nabla ^{i}-i\frac{e}{\hbar c}A_{i}$. Here $\mathbf{%
A}$ is the vector potential.

Further we assume the local density - density interaction Hamiltonian \cite%
{Abrikosov},
\begin{equation}
V=\frac{g^{2}}{2}\int_{\mathbf{r}}\psi _{\alpha }^{+\uparrow }\left( \mathbf{%
r}\right) \psi _{\alpha }^{\downarrow }\left( \mathbf{r}\right) \psi _{\beta
}^{\downarrow +}\left( \mathbf{r}\right) \psi _{\beta }^{\uparrow }\left(
\mathbf{r}\right) \text{,}  \label{int}
\end{equation}%
ignoring the Coulomb repulsion (that as usual is accounted for by a
pseudopotential, so that $g$ is the electron - phonon coupling). It is
important that the interaction has a cutoff Debye frequency $\Omega $, so
that it is active in an energy shell of width $2\hbar \Omega $ around the
Fermi level \cite{Abrikosov}.

\subsection{ Matsubara Green's functions and Gor'kov equations}

Finite temperature properties of the superconducting condensate are
described by the normal and the anomalous Matsubara Green's functions \cite%
{Abrikosov} (GF),%
\begin{eqnarray}
G_{\alpha \beta }^{ts}\left( \mathbf{r}\tau ,\mathbf{r}^{\prime }\tau
^{\prime }\right) &=&-\left \langle T\psi _{\alpha }^{t}\left( \mathbf{r}%
\tau \right) \psi _{\beta }^{\dagger s}\left( \mathbf{r}^{\prime }\tau
^{\prime }\right) \right \rangle ;F_{\alpha \beta }^{ts}\left( \mathbf{r}%
\tau ,\mathbf{r}^{\prime }\tau ^{\prime }\right) =\left \langle T\psi
_{\alpha }^{t}\left( \mathbf{r}\tau \right) \psi _{\beta }^{s}\left( \mathbf{%
r}^{\prime }\tau ^{\prime }\right) \right \rangle ;  \label{GF} \\
F_{\alpha \beta }^{+ts}\left( \mathbf{r}\tau ,\mathbf{r}^{\prime }\tau
^{\prime }\right) &=&\left \langle T\psi _{\alpha }^{\dagger t}\left(
\mathbf{r}\tau \right) \psi _{\beta }^{\dagger s}\left( \mathbf{r}^{\prime
}\tau ^{\prime }\right) \right \rangle ,  \notag
\end{eqnarray}%
with the spin Ansatz
\begin{eqnarray}
G_{\alpha \beta }^{ts}\left( \mathbf{r}\tau ,\mathbf{r}^{\prime }\tau
^{\prime }\right) &=&\delta ^{ts}G_{\alpha \beta }\left( \mathbf{r,r}%
^{\prime },\tau -\tau ^{\prime }\right) ;F_{\alpha \beta }^{ts}\left(
\mathbf{r}\tau ,\mathbf{r}^{\prime }\tau ^{\prime }\right) =-\varepsilon
^{ts}F_{\alpha \beta }\left( \mathbf{r,r}^{\prime },\tau -\tau ^{\prime
}\right) ;  \label{spin Ansatz} \\
F_{\alpha \beta }^{+ts}\left( \mathbf{r}\tau ,\mathbf{r}^{\prime }\tau
^{\prime }\right) &=&\varepsilon ^{ts}F_{\alpha \beta }^{+}\left( \mathbf{r,r%
}^{\prime },\tau -\tau ^{\prime }\right) \text{.}  \notag
\end{eqnarray}%
Here the Plank constant is set to $\hbar =1$. Using the Fourier transform,
\begin{equation}
G_{\gamma \kappa }\left( \mathbf{r},\tau \right) =T\sum \nolimits_{s}\exp %
\left[ -i\omega _{s}\tau \right] G_{\gamma \kappa }\left( \omega ,\mathbf{r}%
\right) \text{,}  \label{Fourier2D}
\end{equation}%
with fermionic Matsubara frequencies, $\omega _{s}=2\pi T\left( s+1/2\right)
$,\ one obtains from equations of operator motion the set of Gor'kov
equations, see \cite{JCM15},\cite{Rosenstein} generalized to include
magnetic field:
\begin{eqnarray}
\left( i\omega +\mu \right) G_{\gamma \kappa }\left( \mathbf{r,r}^{\prime
},\omega \right) +i\ vD_{\mathbf{r}}^{i}\sigma _{\gamma \beta }^{i}G_{\beta
\kappa }\left( \mathbf{r,r}^{\prime },\omega \right) +\Delta _{\alpha \gamma
}\left( \mathbf{r,}0\right) F_{\alpha \kappa }^{+}\left( \mathbf{r,r}%
^{\prime },\omega \right) &=&\delta ^{\gamma \kappa }\delta \left( \mathbf{%
r-r}^{\prime }\right) ;  \notag \\
\left( -i\omega +\mu \right) F_{\gamma \kappa }^{+}\left( \mathbf{r},\mathbf{%
r}^{\prime },\omega \right) -ivD_{\mathbf{r}}^{i}\sigma _{\alpha \gamma
}^{i}F_{\alpha \kappa }^{+}\left( \mathbf{r},\mathbf{r}^{\prime },\omega
\right) -\Delta _{\alpha \gamma }^{\ast }\left( \mathbf{r},0\right)
G_{\alpha \kappa }\left( \mathbf{r},\mathbf{r}^{\prime },\omega \right) &=&0%
\text{.}  \label{GEqua}
\end{eqnarray}%
It will be shown that the singlet pairing pseudo-spin Ansatz, $\Delta
_{\alpha \gamma }\equiv \sigma _{\alpha \gamma }^{x}\Delta $, obeys the
Pauli principle. The gap function consequently reads: $\Delta =\frac{1}{2}Tr%
\left[ \sigma ^{x}\widehat{\Delta }\right] $. Notice, that in contrast to
conventional metals with parabolic dispersion law, in the case of the Weyl
semi - metals the second Gor'kov equation, Eq.(\ref{GEqua}), contains
transposed Pauli matrices for isospins. The applicability of the mean field
approach in a purely 2D model have been widely discussed \cite{RMP} since
(logarithmic) infrared divergences appear in corrections to the
approximation. The corrections of the long range charge density waves
instability are assumed to be cut off by the finite size of the sample, etc.

\section{The transition line in H-T plane}

Near the normal-to-superconducting transition line the gap $\Delta $ is
small and the set of the Gor'kov equations (\ref{GEqua}) can be linearized.
We consider here the 2D case neglecting velocity in magnetic field
direction. More general case will be discussed below. In this case the gap
equation describing the critical curve $H_{c2}\left( T\right) $ has \ the
form, see \cite{Rosenstein} for details,
\begin{eqnarray}
\Delta \left( \mathbf{r}\right)  &=&\frac{g^{2}}{2}T\sum\nolimits_{\omega
}\int_{\mathbf{r}^{\prime }}\Delta ^{\ast }\left( \mathbf{r}^{\prime
}\right) \sigma _{\kappa \beta }^{x}G_{\beta \gamma }^{2}\left( \mathbf{%
r^{\prime },r}\right) \sigma _{\gamma \alpha }^{x}G_{\alpha \kappa
}^{1}\left( \mathbf{r,r}^{\prime }\right)   \label{EqH} \\
&=&g^{2}\sum\nolimits_{\omega }\int_{\mathbf{r}^{\prime }}\Delta ^{\ast
}\left( \mathbf{r}^{\prime }\right) \left(
\begin{array}{c}
G_{22}^{2}\left( \mathbf{r^{\prime },r}\right) G_{1\mathbf{1}}^{1}\left(
\mathbf{r,r}^{\prime }\right) +G_{11}^{2}\left( \mathbf{r^{\prime },r}%
\right) G_{2\mathbf{2}}^{1}\left( \mathbf{r,r}^{\prime }\right)  \\
+G_{12}^{2}\left( \mathbf{r^{\prime },r}\right) G_{12}^{1}\left( \mathbf{r,r}%
^{\prime }\right) +G_{21}^{2}\left( \mathbf{r^{\prime },r}\right) G_{2%
\mathbf{1}}^{1}\left( \mathbf{r,r}^{\prime }\right)
\end{array}%
\right) \text{.}  \notag
\end{eqnarray}%
Here the normal GF is obtained from,$\ $%
\begin{equation}
\left[ \ iv\mathbf{D}_{\mathbf{r}}\cdot \mathbf{\sigma }_{\gamma \beta
}+\left( i\omega +\mu \right) \delta _{\gamma \beta }\right] G_{\beta \kappa
}^{1}\left( \mathbf{r,r}^{\prime }\right) =\delta ^{\gamma \kappa }\delta
\left( \mathbf{r-r}^{\prime }\right) \text{,}  \label{normalGF}
\end{equation}%
while a quantity$\ G_{\beta \gamma }^{2}$ (an auxiliary function associated
with $G$ via a product of an axis reflection and time reversal) obeys a
\textit{different} equations:
\begin{equation}
\left[ -iv\mathbf{D}_{\mathbf{r}}\cdot \mathbf{\sigma }_{\gamma \beta
}^{t}+\left( -i\omega +\mu \right) \delta _{\gamma \beta }\right] G_{\beta
\kappa }^{2}\left( \mathbf{r}^{\prime }\mathbf{,r}\right) =\delta ^{\gamma
\kappa }\delta \left( \mathbf{r-r}^{\prime }\right) \text{.}  \label{n2a}
\end{equation}%
Here $\mathbf{\sigma }^{t}$ is the transposed Pauli matrix that replaces $%
\mathbf{\sigma }$ in the customary normal state Weyl equation for left
movers, Eq.(\ref{normalGF}).

In the uniform magnetic field the GF can be written in the symmetric gauge, $%
\mathbf{A=}\frac{1}{2}\mathbf{H\times r}$, in the following form:%
\begin{eqnarray}
G_{\beta \kappa }^{1}\left( \mathbf{r},\mathbf{r}^{\prime }\right) &=&\exp %
\left[ -i\frac{xy^{\prime }-yx^{\prime }}{2l^{2}}\right] g_{\beta \kappa
}^{1}\left( \mathbf{r-r}^{\prime }\right) ;  \label{n3a} \\
G_{\beta \kappa }^{2}\left( \mathbf{r}^{\prime },\mathbf{r}\right) &=&\exp %
\left[ -i\frac{xy^{\prime }-yx^{\prime }}{2l^{2}}\right] g_{\beta \kappa
}^{2}\left( \mathbf{r}^{\prime }\mathbf{-r}\right) \text{.}  \notag
\end{eqnarray}%
Here $l^{2}=c/eH$ is the magnetic length. This phase Ansatz indeed works.
Substituting it into Eq.(\ref{normalGF}) and Eq.(\ref{n2a}) respectively,
the variables separate. It reads,
\begin{equation}
L_{\gamma \beta }^{1}g_{\beta \kappa }^{1}\left( \mathbf{r-r}^{\prime
}\right) =\delta ^{\gamma \kappa }\delta \left( \mathbf{r-r}^{\prime
}\right) ;L_{\gamma \beta }^{2}g_{0\beta \kappa }^{2}\left( \mathbf{r-r}%
^{\prime }\right) =\delta ^{\gamma \kappa }\delta \left( \mathbf{r-r}%
^{\prime }\right) \text{,}  \label{n3aa}
\end{equation}%
where $L_{\gamma \beta }^{1}=\left[ \left( i\omega +\mu \right) \delta
_{\gamma \beta }+\left( -iv\sigma _{\gamma \beta }^{i}\ \nabla
_{r}^{i}\right) \right] $ , $L_{\gamma \beta }^{2}=\left[ \left( -i\omega
+\mu \right) \delta _{\gamma \beta }-iv\sigma _{\gamma \beta }^{it}\nabla
_{r}^{i}\right] $.

Solutions of these equations gives;

\begin{eqnarray}
g_{22}^{1}\left( \mathbf{p}\right) &=&z^{\ast -1}\left( i\omega +\mu \right)
;\text{ \ }g_{12}^{1}\left( \mathbf{p}\right) =-z^{\ast -1}v\left(
p_{x}-ip_{y}\right) ;g_{11}^{1}\left( \mathbf{p}\right) =z^{\ast -1}\left(
i\omega +\mu \right) ;  \label{n4a} \\
\text{ }g_{21}^{1}\left( \mathbf{p}\right) &=&-z^{\ast -1}v\left(
p_{x}+ip_{y}\right) ;g_{11}^{2}\left( \mathbf{p}\right) \ =z^{-1}\ \left(
-i\omega +\mu \right) ;\text{ }g_{12}^{2}\left( \mathbf{p}\right)
=-z^{-1}v\left( p_{x}+ip_{y}\right) ;  \notag \\
g_{22}^{2}\left( \mathbf{p}\right) &=&z^{-1}\left( -i\omega +\mu \right) ;%
\text{ \ }g_{21}^{2}\left( \mathbf{p}\right) =-z^{-1}v\left(
p_{x}-ip_{y}\right) \ ;  \notag
\end{eqnarray}%
where $z=\left( -i\omega +\mu \right) ^{2}-\left( vp\right) ^{2}$ and $p$
denote quasi-momentum.

Substituting the phase factors of GF from Eq.(\ref{n3a}) into the gap
equation, Eq.(\ref{EqH}), one obtains:

\begin{equation}
\Delta \left( \mathbf{r}\right) =g^{2}T\sum\nolimits_{\omega }\int_{\mathbf{r%
}^{\prime }}\exp \left[ \ i\frac{\left[ \mathbf{r\times \rho }\right] }{l^{2}%
}\right] \Delta ^{\ast }\left( \mathbf{r}^{\prime }\right) \left(
\begin{array}{c}
g_{22}^{2}\left( \mathbf{-\rho }\right) g_{11}^{1}\left( \mathbf{\rho }%
\right) +g_{11}^{2}\left( \mathbf{-\rho }\right) g_{22}^{1}\left( \mathbf{%
\rho }\right)  \\
+g_{12}^{2}\left( \mathbf{-\rho }\right) g_{12}^{1}\left( \mathbf{\rho }%
\right) +g_{21}^{2}\left( \mathbf{-\rho }\right) g_{21}^{1}\left( \mathbf{%
\rho }\right)
\end{array}%
\right) ,  \label{n5a}
\end{equation}%
where $\mathbf{\rho }\mathbf{=}\mathbf{r-r}^{\prime }$. Looking for the gap
function $\Delta \left( \mathbf{r}\right) $ in the form\cite{India} $\Delta
\left( \mathbf{r}\right) =\Delta \exp \left( -\frac{r^{2}}{2l^{2}}\right) $,
one obtains after Fourier transformation.
\begin{equation}
1=g^{2}T\sum\nolimits_{\omega }\int \mathbf{d}^{2}\mathbf{\rho }\exp \left[
\ i\frac{\left[ \mathbf{r\times \rho }\right] -\mathbf{r\rho }}{l^{2}}\right]
\exp \left( -\frac{\mathbf{\rho }^{2}}{2l^{2}}\right) \int \frac{d^{2}%
\mathbf{p}d^{2}\mathbf{q}}{\left( 2\pi \right) ^{4}}e^{i\mathbf{q\rho }%
}S\left( \mathbf{p,q}\right) ,  \label{n5aa}
\end{equation}%
where
\begin{equation}
S\left( \mathbf{p,q}\right) =\frac{2\left[ \omega ^{2}+\mu ^{2}+v^{2}\left(
\mathbf{p-q}\right) \cdot \mathbf{p}\right] }{\left[ \left( -i\omega +\mu
\right) ^{2}-v^{2}\left( \mathbf{p-q}\right) ^{2}\right] \left[ \left(
i\omega +\mu \right) ^{2}-v^{2}\mathbf{p}^{2}\right] }  \label{n6aa}
\end{equation}%
In the polar coordinates Eq.\ref{n6aa} has the form

\begin{eqnarray}
1 &=&\frac{2g^{2}T}{v^{4}\left( 2\pi \right) ^{2}}\sum\nolimits_{\omega
}\int \rho d\rho d\Theta d\varepsilon _{q}\ d\varepsilon _{p}d\psi \exp %
\left[ \ \frac{ir\rho \sin \Theta -r\rho \cos \Theta }{l^{2}}-\frac{\rho ^{2}%
}{2l^{2}}\right] \cdot   \label{n7aa} \\
&&\cdot J_{0}\left( \frac{\varepsilon _{q}\rho }{v}\right) \varepsilon
_{q}\varepsilon _{p}\frac{\omega ^{2}+\mu ^{2}+\varepsilon _{p}^{2}\mathbf{-}%
\varepsilon _{p}\varepsilon _{q}\cos \psi }{\left[ \left( -i\omega +\mu
\right) ^{2}-\left( \varepsilon _{p}^{2}-2\varepsilon _{p}\varepsilon
_{q}\cos \psi +\varepsilon _{q}^{2}\right) \right] \left[ \left( i\omega
+\mu \right) ^{2}-\varepsilon _{p}^{2}\right] }.\   \notag
\end{eqnarray}%
Here $\Theta $ is the angle between vectors $\mathbf{r}$ and $\mathbf{\rho }$
while $\psi $ is the angle between momentum $\mathbf{p}$ and $\mathbf{q}$, $%
vp=\varepsilon _{p};vq=\varepsilon _{q},$ $J_{0}(z)$ is the Bessel function.

Usually within the BCS approach, the interaction is approximated not just by
a contact in space and a step function - like cutoff,%
\begin{equation}
\mu -\hbar \Omega <\varepsilon _{p}<\mu +\hbar \Omega \text{,}  \label{shell}
\end{equation}%
Therefore the in Eq.(\ref{n8aaa}) is restricted. This is unphysical since
the step function dependence is just an approximation of a more realistic
second order effective electron interaction due to phonon exchange.

Neglecting the dispersion of the optical phonon, the sharp cutoff will be
replaced by the Lorenzian function
\begin{equation}
f\left( \varepsilon _{p}\right) =\frac{\Omega ^{2}}{\Omega ^{2}+\left(
\varepsilon _{p}-\mu \right) ^{2}}\text{.}  \label{fn}
\end{equation}

Performing integral \cite{Gradshtein},\cite{koshelev} over angle $\Theta $
and $\rho $ one obtains the coexistence curve at $H-T$ plane in reduced
variables with barred energies denoting division by temperature $T$: $\left(
\overline{\mu }=\mu /T,\overline{\omega }=\omega /T,\overline{\varepsilon }%
_{p,q}=\varepsilon _{p.q}/T\right) $

\begin{equation}
\frac{1}{\lambda }=\frac{2\chi }{\overline{\mu }}\sum\nolimits_{\omega }\int
d\overline{\varepsilon }_{q}d\overline{\varepsilon }_{p}d\psi \frac{f\left(
\overline{\varepsilon }_{p}\right) \overline{\varepsilon }_{q}\overline{%
\varepsilon }_{p}\exp \left[ -\chi \overline{\varepsilon }_{q}^{2}\right]
\left( \overline{\omega }^{2}+\overline{\mu }^{2}+\overline{\varepsilon }%
_{p}^{2}\mathbf{-}\overline{\varepsilon }_{q}\overline{\varepsilon }_{p}\cos
\psi \right) }{\left[ \left( -i\overline{\omega }+\overline{\mu }\right)
^{2}-\left( \overline{\varepsilon }_{p}^{2}+\overline{\varepsilon }_{q}^{2}-2%
\overline{\varepsilon }_{p}\overline{\varepsilon }_{q}\cos \psi \right) %
\right] \left[ \left( i\overline{\omega }+\overline{\mu }\right) ^{2}-%
\overline{\varepsilon }_{p}^{2}\right] }\text{.}  \label{n8aaa}
\end{equation}%
Here $\lambda $ $=D\left( \mu \right) g^{2}$ is the electron-electron
strength constant, $D\left( \mu \right) =\mu /4\pi v^{2}$ is the density of
free $2D$ Dirac electron gas. The dimensionless magnetic field parameter in
exponent is defined by $\chi =c\sqrt{\hbar /v}/2et^{2}h,$ where $t\equiv
T/\Omega ,t_{c}=T_{c}/\Omega $, Performing summation over Matsubara
frequencies and numerical integrations, one obtain the upper critical
magnetic fields depending on temperature and adiabaticity parameter $\gamma
=\mu /\left( \hbar \Omega \right) $.

\begin{figure}[tbp]
\begin{center}
\includegraphics[width=12cm]{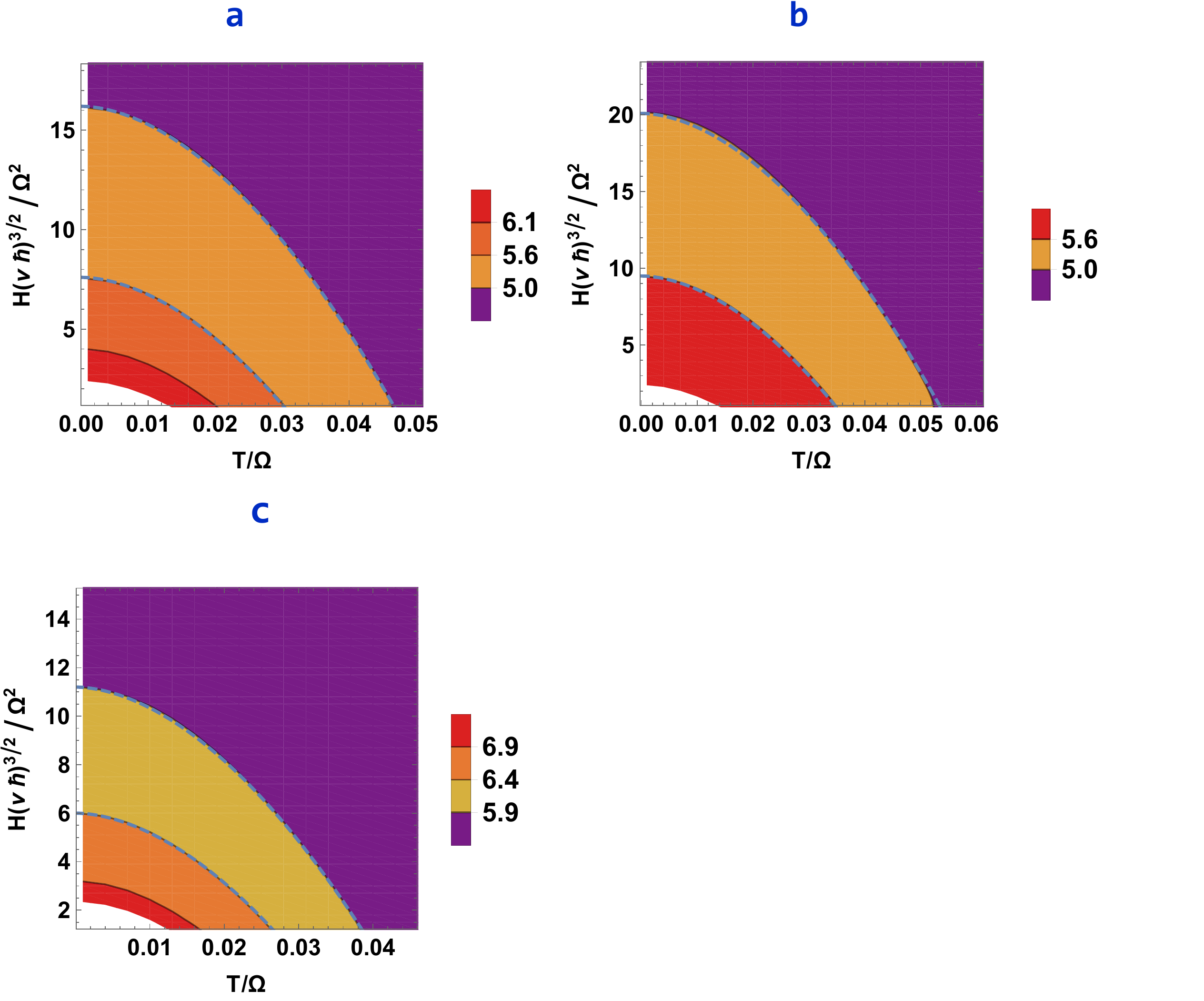}
\end{center}
\par
\vspace{-0.5cm}
\caption{Upper critical field as a function on temperature for various
adiabatic ratio $\protect\gamma $ and electron-electron strength values $%
\protect\lambda $ a) $\protect\gamma =10$ b) \ $\protect\gamma =3$ c) $%
\protect\gamma =1.$ Dash lines described by the universal interpolation
formula (\protect\ref{IF}) with reduced magnetic field $h^{\ast }=0.55$
parameter. Color scale marks $\protect\lambda ^{-1}$ magnitudes.}
\end{figure}

In Fig. 1 magnetic field is given in units of magnetic field $H_{u}=\Omega
^{2}/\left( v\hbar \right) ^{3/2}$ for three different values of the
adiabaticity parameter $\gamma $ and the DSM superconductors electron
velocity $v=10^{7}cm/\sec ,$ typical to these materials. The temperature
dependence $H_{c2}\left( T\right) $ is fitted very well by an interpolation
formula,%
\begin{equation}
H\left( T\right) =H\left( 0\right) \left( 1-\left( T/T_{c}\right)
^{1/h^{\ast }}\right) \text{,}  \label{IF}
\end{equation}%
for the universal value of the reduced magnetic field $h^{\ast }=0.55$
determining the exponent. Results of the theory demonstrates excellent
agreement with temperature dependence of the upper critical magnetic field
measured in DSM $TaP$ \cite{TaP}, see Fig.2.
\begin{figure}[tbp]
\begin{center}
\includegraphics[width=12cm]{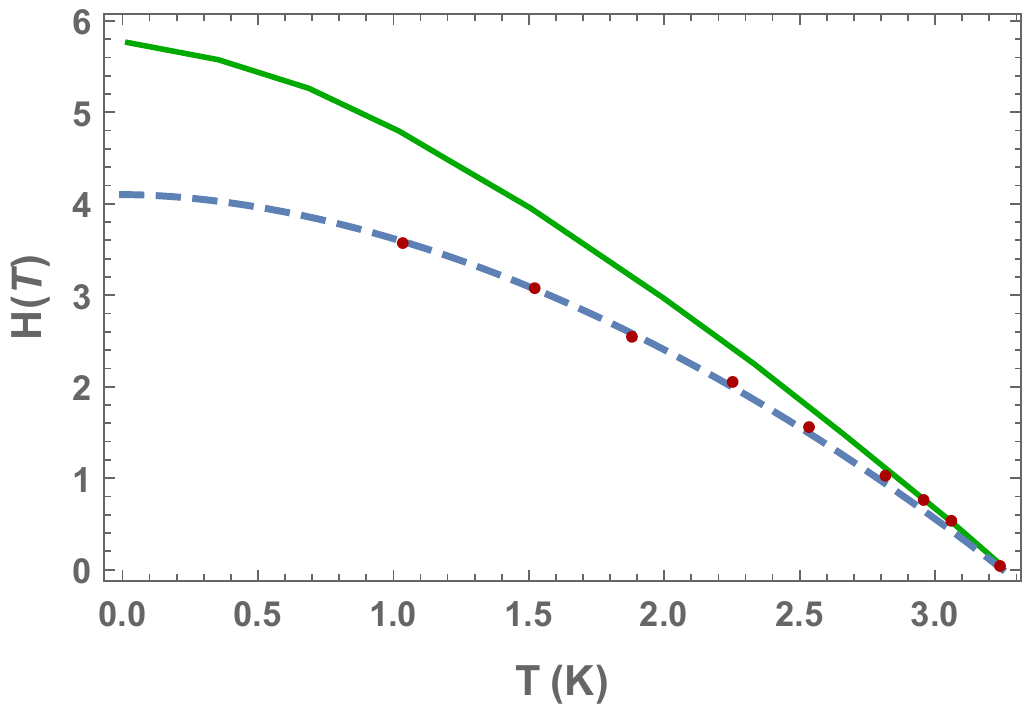}
\end{center}
\par
\vspace{-0.5cm}
\caption{Upper Critical magnetic field $H_{c2}$ versus temperature for DSM $%
TaP$. The green curve marks result of the WHH\ theory for clean conventional
superconductor. Red dots are the experimental data while the dash line is
result of our theory.}
\end{figure}

In more general case when velocity in the $z$ axis direction (parallel to
the magnetic field) $v_{z}$ is not zero, the upper critical field does not
depends on $v_{z}$ since it renormalized the density of states and hence the
electron-electron strength constant. The reduced magnetic field $h^{\ast }$
in this case coincides with that calculated for layered system (similar to
conventional superconductors \cite{Bulaevsky}).

\section{Conclusions}

In this paper,the microscopic semi-classical theory of phonon mediated
superconductivity in Dirac semimetals at magnetic fields was constructed in
entire range of the temperature. Main results are presented in Figs. 1-2.
Within the weak coupling approach, the retardation effects were explicitly
taken into account by the dispersionless model of the electron - phonon
coupling, Eq.(\ref{fn}). This is of importance since commonly used step
function produce spurious oscillation\cite{RMP,strongfield}. The upper
critical magnetic field $H_{c2}$ for graphene like Dirac semimetal was
calculated for all temperatures in 2D and 3D Dirac superconductors. It is
predicted that in Dirac semimetals the upper critical magnetic field $%
H_{c2}\left( T\right) $ is lower than predicted by the conventional
Werthamer -Helfand -Hohenberg formula \cite{WHH} derived for superconductors
with parabolic dispersion relation. The reduced magnetic field ratio is
universal (independent of the chemical potential, interaction strength etc.)
and smaller than the Werthamer ratio $h^{\ast }=H_{c2}\left( 0\right)
/\left( -T\frac{dH_{c2}}{dT}\right) |_{T_{c}}$ for clean superconductors: $%
h^{\ast }=0.55$ for WSM, $h^{\ast }\left( 0\right) =0.69$ for parabolic
band. This explains the recent experiments on $Cd_{3}As_{2},ZrTe_{5}$ and
especially on $TaP$ (see ref. \cite{CdAs}, \cite{superWSM3D} and \cite{TaP}
respectively). Going beyond semi - classical approximation is typically more
complicated and has been contemplated in parabolic band materials \cite%
{India} and recently in Weyl semimetals\cite{strongfield,Manivnew}.

\textit{Acknowledgements.}

We are grateful to T. Maniv for valuable discussions.\ B. R. acknowledges
MOST of ROC grant 107-2112-M-009-009-MY3, hospitality of Peking and Bar Ilan
Universities. D.P. Li was supported by National Natural Science Foundation
of China (Nos. 11274018 and 11674007).

\end{document}